\documentclass{article}

\usepackage{arxiv}

\usepackage[utf8]{inputenc} 
\usepackage[T1]{fontenc}    
\usepackage{hyperref}       
\usepackage{url}            
\usepackage{booktabs}       
\usepackage{amsfonts}       
\usepackage{nicefrac}       
\usepackage{microtype}      
\usepackage{subfigure}

\usepackage{graphicx}
\graphicspath{ {./} }
\usepackage{amsmath}
\usepackage{graphicx}
\usepackage{amsmath}
\usepackage{color,soul}

\usepackage{tikz}
\usepackage[english]{babel} 
\usepackage{authblk}

\newtheorem{statement}{Statement}

\begin{document}
\title{Existence conditions for hidden feedback loops in online recommender systems}
\renewcommand{\undertitle}{A preprint}
\renewcommand{\shorttitle}{Exist. cond. for hidd. feedback loops in recom. sys.}

\author{Anton S.~Khritankov} 
\author{Anton A.~Pilkevich}

\affil{Moscow Institute of Physics and Technology \\
        Dolgoprudny, Moscow Region, \\
        Russian Federation \\
        \texttt{anton.khritankov@phystech.edu}}

\hypersetup{
pdftitle={Existence conditions for hidden feedback loops in online recommender systems},
pdfsubject={cs.LG},
pdfauthor={Anton S.~Khritankov, Anton A.~Pilkevich},
pdfkeywords={hidden feedback loops, recommender systems, multi-armed bandits},
}

\maketitle             
\begin{abstract}
We explore a hidden feedback loops effect in online recommender systems. Feedback loops result in degradation of online multi-armed bandit (MAB) recommendations to a small subset and loss of coverage and novelty. We study how uncertainty and noise in user interests influence the existence of feedback loops. 
First, we show that an unbiased additive random noise in user interests does not prevent a feedback loop. Second, we demonstrate that a non-zero probability of resetting user interests is sufficient to limit the feedback loop and estimate the size of the effect. 
Our experiments confirm the theoretical findings in a simulated environment for four bandit algorithms.

\end{abstract}
\section{Introduction}

As research and applications of machine intelligence progress, more concerns are arising questioning whether the deployed AI systems fair, explainable, and ethical. Recommender systems are ubiquitous in social networks, streaming services, e-commerce, web marketing and advertising, web-search to provide personalized experience. Their algorithms use the new data from the system to improve their future performance. A positive feedback loop results in that only a small subset of available items is presented to the user. This effect is observed when user interests are being reinforced by previous exposure to specific items or item categories, thus producing self-induced concept drift.

The feedback loop effect is studied in real and model systems as an undesirable phenomenon related to reliable and ethical AI. Some of the consequences of the effect are induced shift in users interests \cite{jiang2019degenerate}, loss of novelty and diversity in recommendations \cite{ziarani2021serendipity}, presence of "echo chambers" and "filter bubbles" \cite{chamber,bubble2}, induced concept drift in housing prices prediction \cite{Khritankov2021Hidden}. Nevertheless, a full description of the feedback loop effect and its existence conditions is still lacking for many cases \cite{krueger2020hidden,adam2020hidden}.

In this paper we extend prior results \cite{jiang2019degenerate} and explore existence conditions of feedback loops in presence of noise in the user behavior. We specifically consider noise as a random shift in user interests to items and categories the user is exposed to. In Section \ref{sect:statement} we theoretically derive existence conditions for two noise models and explore our predictions empirically in sections \ref{sect:experiment} and \ref{sect:analysis} with four bandit algorithms in a simulated environment. 

\section{Related work}

Multi-armed bandits are commonly used in online recommendation systems and experiment design \cite{karimi2018news,burtini2015survey}. Earlier results demonstrate how feedback loops influence distribution of recommendations and user preferences. In \cite{riquelme2018deep} authors show that posterior distribution in Thompson Sampling algorithms is affected by the feedback loop, which worsens regret performance.

In \cite{jiang2019degenerate} authors argue that a feedback loop would exist in a stochastic multi-armed bandit recommender system under some mild assumptions unless a set of available items does not grow at least linearly. Whereas noise and sudden shifts in user interest are important phenomena that occur in practice, they are not usually taken into account. An important contribution of this paper is that we show that a feedback loop would not occur even with a bounded set of available items if user interests are affected by specific kinds of random noise.

\section{Problem statement}
\label{sect:statement}

\subsection{A recommender system model}
\label{sect:system}

Let us consider a recommender system with a single user and available items $\mathbb{M} = \{1,..,M\}$. At time step $t$ the system selects different items $A_t = (a_t^1,.., a_t^l)$ from the set of available items $l < M$ and presents them to the user. User interest to item $a_i$ at step $t$ is described by a  function $\mu_t: \mathbb{M} \to \mathbb{R}^M$. Larger values of $\mu_t(a^i)$ correspond to stronger interest. After that the user examines the items and responds with $c_t = (c_t^1,..,c_t^l)$, $c_t^i \in \{0, 1\}$ sampled independently proportionally to the user interest in item $\mu_t(a^i)$. Therefore, we can model the response at step $t$ with a random variable that has Bernoulli distribution: 
$ c_t^i \sim Bern (\sigma(\mu_t(a_t^i))), \sigma(x) = 1 / (1+e^{-x})$. 
In the simple case \cite{jiang2019degenerate}, the evolution of user interests abides to the monotonicity constraint for each element $a^i$ with respect to $t$ and user interest to item $a_i$ is updated according to the rule:
\begin{equation}
\label{eq:mu-update}
    \mu_{t+1}^i - \mu_t^i = 
    \begin{cases}
       \delta_t, \text{ when } a_i \in A_t, c_t = 1, \\
       - \delta_t, \text{ when } a_i \in A_t, c_t = 0, \\
       0, \text{ when } a_i \notin A_t, \\
    \end{cases}
\end{equation}
where $\delta_t \sim \text{Uniform}[0, 0.01]$ indicates how much does the user interest change at step $t$. The initial user interest $\mu_0^i$ to item $a_i$ is a random variable with a uniform distribution $\mu_0^i \sim \text{Uniform}[-1, 1]$. A single user assumption is justified when users act independently and recommendations for one user do not affect others.  

Following Jiang et al. \cite{jiang2019degenerate} we define a \emph{positive feedback loop} as a situation in the behavior of a recommender system when a $2-$norm of user interests grows to infinity with step number $t$:
\begin{equation}
\label{eq:loop}
\lim_{t \to \infty} \|\mu_t - \mu_0 \|_2 = \infty. 
\end{equation}

\subsection{Multi-armed bandit problem statement}
\label{sect:bandits}

Let us sate an online MAB recommendation problem. There is a set of $M$ levers and an agent. Each lever has a probability distribution of reward associated with it. The agent can play $l < M$ levers and get rewards from each of the levers. The goal of the agent is to maximize the reward or, in other terms, minimize the regret, which is difference between the achieved and the maximum total reward. We correspond levers with available items $a_i$ and rewards with user responses $c_t^i$ at step $t$. User is the environment and the agent executes an item selection policy $S$. The optimization problem that the agent is trying to solve is $T \cdot l - \sum_{t = 1}^T \sum_{i = 1}^l c_t^i \to \min_{S}$. We study the following bandit algorithms.

\paragraph{Thompson Sampling}\cite{slivkins2019introduction}. We define Bernoulli random variables $\pi_t(\theta_1),.., \pi_t(\theta_M)$ that correspond to the winning $c_t^i = 1$ probability if a lever is played and initialize them at $t=0$ via prior distribution $\theta_i \sim Beta(1, 1) = \text{Uniform}[0, 1]$. The posterior distribution conditioned on user responses is then given by $Beta(\alpha_t^i, \beta_t^i)$ for each $a^i \in \mathbb{M}$. Distribution parameters $\alpha_t^i, \beta_t^i$ are updated for each $a_i \in A_t$ based on user response as 
\begin{equation}\label{eq:ts-update}
    \begin{gathered}
        \alpha_{t+1}^i = \alpha_t^i + c_t^i, \beta_{t+1}^i = \beta_t^i + 1 - c_t^i.
    \end{gathered}
\end{equation}

\paragraph{$\epsilon$-greedy}\cite{slivkins2019introduction}. 
With probability $\epsilon$ the policy decides to explore and selects $l$ levers at random uniformly. With probability $1 - \epsilon$ it returns top $l$ levers with the highest mean rewards $d_t^i / n_t^i$, $d_t^i$ is total accumulated rewards for item $a_i$. Policy state is updated from the user feedback $c_t^i \in \{0, 1\}$ as follows:

\begin{equation}
\label{eq:greedy-update}
\begin{gathered}
n_{t+1}^i = n_t^i + \mathbb{I}\{a_i \in A_t\}, d_{t+1}^i = d_{t}^i + c_t^i.
\end{gathered}
\end{equation}

\paragraph{Optimal.} The policy selects $l$ items with the highest user interest $\mu_t^i$ at each step. There is no internal state for the policy that needs to be updated. The policy is optimal in a sense that it knows the actual user interests and selects levers with the highest expected reward.

\paragraph{Random.} The policy selects each item from $M$ with the same probability $1/|M|$ at random. The policy does not have any internal state to update.

\subsection{Additive noise model}
\label{sect:add-noise-model}

Let us drop the assumption made in Section \ref{sect:statement} that user interests $\mu_t^i$ is a known real vector with a stochastic update rule. Instead we model user interest in item $a_i$ as a random function with known mean $\mu_t^i = \bar{\mu}_t^i +  \omega_t^i$, where $\omega_t^i$ is an unbiased random noise, $\mathbf{E}\, \omega_t^i = 0$ and $\bar{\mu}_t^i = \mathbf{E}\, \mu_t^i$. The interest update rule \eqref{eq:mu-update} then becomes:
\begin{equation}
\label{eq:add-noise}
    \begin{gathered}
        \bar{\mu}_{t+1}^i - \bar{\mu}_{t}^i = \delta_t c_t^i - \delta_t (1 - c_t^i), \text{ if } i \in A_t \\
        \bar{\mu}_{t+1}^i - \bar{\mu}_{t}^i = 0 \text{ otherwise},
    \end{gathered}
\end{equation}
where $\delta_t$ remains the same as in \eqref{eq:mu-update} and $c_t^i$ now depends on the noisy user interest $\mu_t^i$.
We consider a case with uniform and bounded noise $\omega_t^i \sim \text{Uniform}[-w, w], w > 0$. Then the response $c_t^i \sim Bern \left(\sigma(\bar{\mu}_t^i + \omega_t^i) \right)$. For further analysis we define a \emph{constant best levers} condition when for all $t > t^0$ the set of selected arms remains the same $A_t = A_{t^0}$.

\begin{statement}
\label{stmt:add-noise}
Let the recommender system with noise \eqref{eq:add-noise} satisfy the constant best levers condition. Then 
\begin{equation}
    \label{eq:add-existence}
    \lim_{t \to \infty} \|\mu_t^i - \mu_0^i \|_2 = \infty, \forall w \geq 0, a_i \in A_{t^0}
\end{equation}
\end{statement}

As follows from this statement any bounded unbiased additive noise in a form of [\ref{eq:add-noise}] does not prevent a feedback loop from occurring.

\subsection{Interest restarts model}

Users may sometimes lose or forget their interest to items. We call such event an \emph{interest restart}. Restarts may be caused by satisfaction of the interest, change of users agenda outside of the system or introduction of a competing interest, disappearance of the item itself and other reasons. We consider a linear model of the aforementioned effect. When a restart occurs with probability $q$, then user interest to item $a_i$ is replaced with a new random value that follows the same uniform initial distribution as $\mu_0^i$, or is reduced by $0 \leq s \leq 1$:

\begin{equation}
\label{eq:restarts}
    \mu_{t+1}^i = 
    \begin{cases}
      \mu_{t}^i + \Delta_t^i, \text{ with probability } 1 - q, \\
      (1 - s) \mu_0 + s (\mu_t^i + \Delta_t^i), \text{ otherwise},
    \end{cases}
\end{equation}
where $\Delta_t^i = \mu_{t+1}^i - \mu_t^i$ as defined in (\ref{eq:mu-update}). Note that now the update rule is not monotonic.

\begin{statement}
\label{stmt:restarts}
Let the recommender system with restarts \eqref{eq:restarts},\eqref{eq:mu-update} satisfy the constant best levers condition. If $\mu_0^i > 0$ then the expected user interest is bounded by
\begin{equation}
\label{eq:restarts-cond}
    \mathbf{E}\, \delta_t \left( \frac{1}{(1 - s) q} - 1 \right) > \mathbf{E}\,\mu_t^i \gg 0.
\end{equation}
\end{statement}

\section{Experiment}
\label{sect:experiment}

\subsection{Experiment design}

We set up an empirical evaluation to test if our assumptions about system behavior are valid and to check whether the theoretical results hold in practical conditions. \emph{RQ 1.} A positive feedback loop occurs even when an unbiased additive noise is added to user interests as demonstrated by Statement \ref{stmt:add-noise}. \emph{RQ 2.} The system does not need to exhibit exactly the constant best levers regime for results of Statement \ref{stmt:add-noise} and Statement \ref{stmt:restarts} to hold. \emph{RQ 3.} Results of Statement \ref{stmt:add-noise} and Statement \ref{stmt:restarts} hold for a range of different lever selection polices.

We implement a recommender system \cite{khritankov2021banditrepo} described in Section \ref{sect:system} and compare behavior different lever selection policies. We include Thomson Sampling (TS) and $\epsilon$-greedy trainable selection policies, and Random and Optimal policies as baselines. Both latter policies do not require learning. The following parameters are varied during the experiment: number of items available for recommendation $M$, how many items are selected $l$, each of the algorithms parameters. We run the experiment for a maximum of $T$ steps and repeat it with the same parameters in several trials to get an estimate of the confidence interval for results. For TS policy the priors are set to $\alpha_0^i = 1, \beta_0^i = 1, i = \{1,..,M\}$ to reflect the uniform distribution. The policy samples rewards from the prior $r_i \sim Beta(\alpha_t^i, \beta_t^i)$, selects $l$ items with the highest sampled rewards $A_t$ and after that updates the priors based on user feedback \eqref{eq:ts-update}. The Random and Optimal policies do not require initialization. The Optimal policy returns $l$ items with the highest interests $A_t$ and is updated with new user interests $\bar{\mu}_t^i$ at each simulation step. The $\epsilon$-greedy policy takes $\epsilon$ as a constant parameter set before the trial starts, with probability $\epsilon$ selects items at random uniformly or returns $l$ items with maximum accumulated rewards. The policy is updated with user responses $c_t$ at each step \eqref{eq:greedy-update}. For Additive noise model we set the amount of noise $w$ parameter. In Interest restarts model we set a probability of restart $q$ and scale $s$ parameters.

The experiment proceeds as follows. A grid of experiment parameters with with $l \leq M$ (see the experiment specification) is set up before start and random seed is fixed. Then tuples of parameters are retrieved from the grid and a number of trials is run. At each trial an experiment instance is initialized with parameters taken from the tuple. The selected user interests model and policy are initialized with corresponding parameters. Only one of the models and policies are used in each trial. At each step we store user feedback $c_t$, interests $\mu_t^i$ and the state of the policy: ${\alpha_t^i, \beta_t^i}$ for TS policy, $d_t^i$ for $\epsilon$-greedy policy. 

\subsection{Experiment results}

\begin{figure}[t]
    \centering
    \includegraphics[scale=0.35, trim=15 15 0 15, clip]{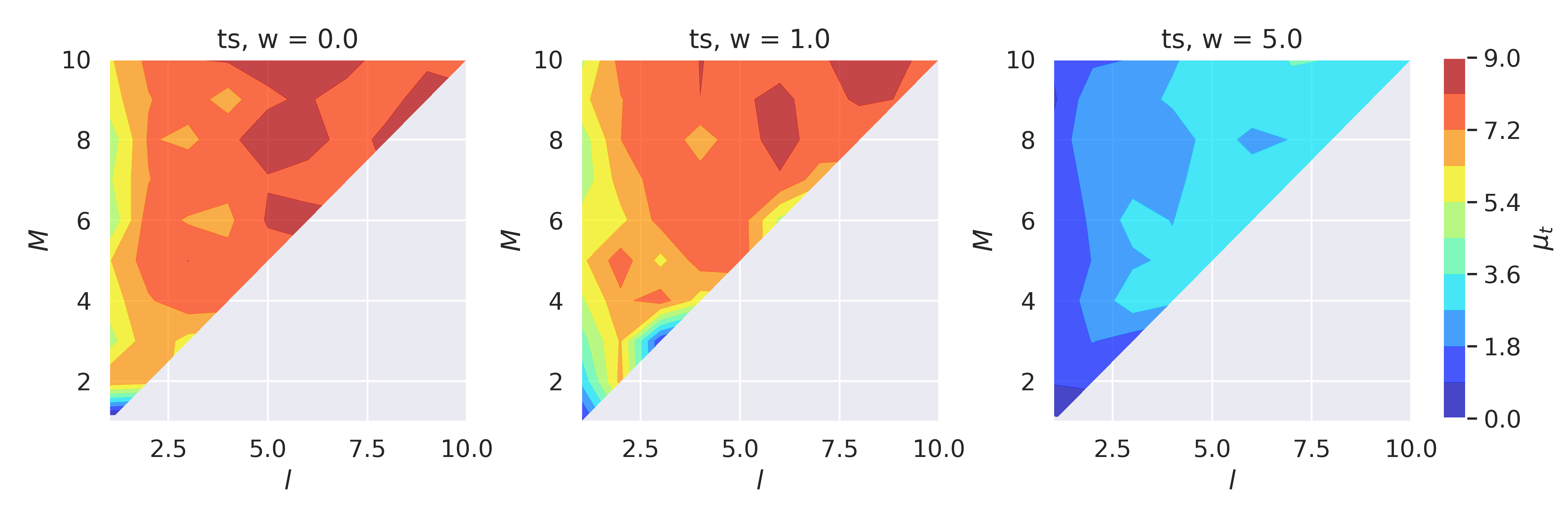}
    \caption{Maximum user interest (color) for Thompson Sampling (TS) policy with different $M$ and $l$ (y- and x- axes) and strength of additive noise $w$ at step $t=2000$.}
    \label{fig:add-noise-ts}
\end{figure}

We show how the total reward changes for the Thompson Sampling (TS) selection policy when the step number $0 \leq t \leq 2000$ at Fig.\,\ref{fig:add-noise-ts} with different values of additive noise $w \in \{0.0, 0.3, 1.0, 3.0, 5.0, 10.0\}$. Note that the figure uses a logarithmic scale. The maximum value of user interest $\max_i \mu_t^i$ and thus amplitude of the feedback loop $\|\mu_t^i - \mu_0^i\|_2$ grows for all $w$ and this does not contradict Statement\,\ref{stmt:add-noise}. Results are averaged and confidence intervals are estimated over 30 runs. Although the growth rate decreases with $w$ for all values of $l < M$ considered in the experiment. Thus for the parameters explored in the experiment we can confirm RQ\,1 for TS selection policy.

\begin{figure}[t]
    \centering
    \includegraphics[scale=0.33, trim=15 15 10 15, clip]{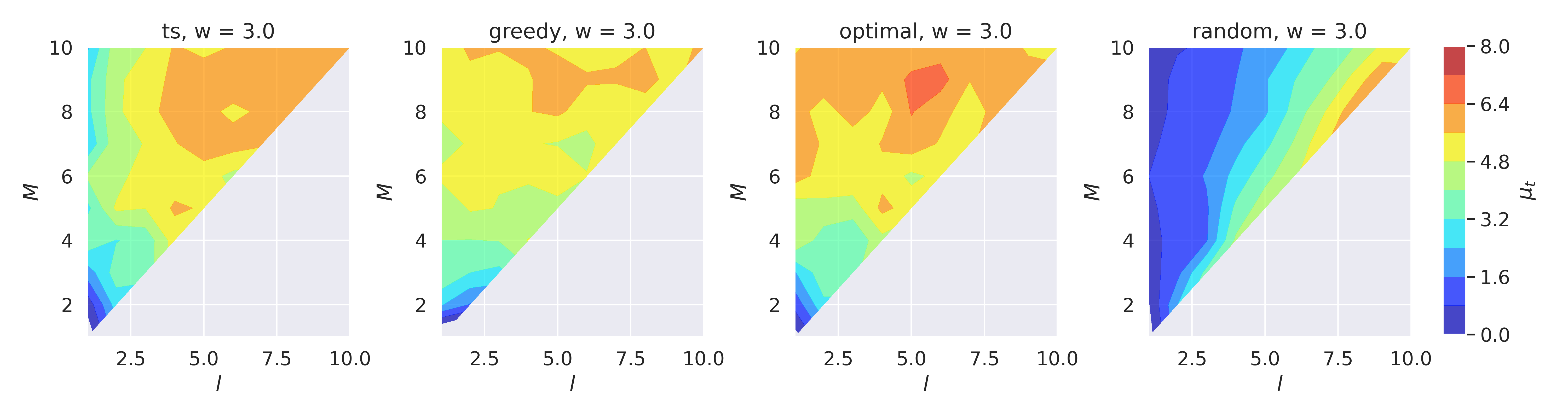}
    \caption{Maximum user interest (color) for Thompson Sampling (ts), Random, Optimal and $\epsilon$-greedy (greedy) selection policies. Axes: y-available items $M$, x- selected items $l$. Total number of steps $T=2000$, strength of additive noise $w=3.0$ }
    \label{fig:add-noise-policies}
\end{figure}

\begin{figure}[t]
    \centering
    \includegraphics[scale=0.39, trim=125 420 70 410, clip]{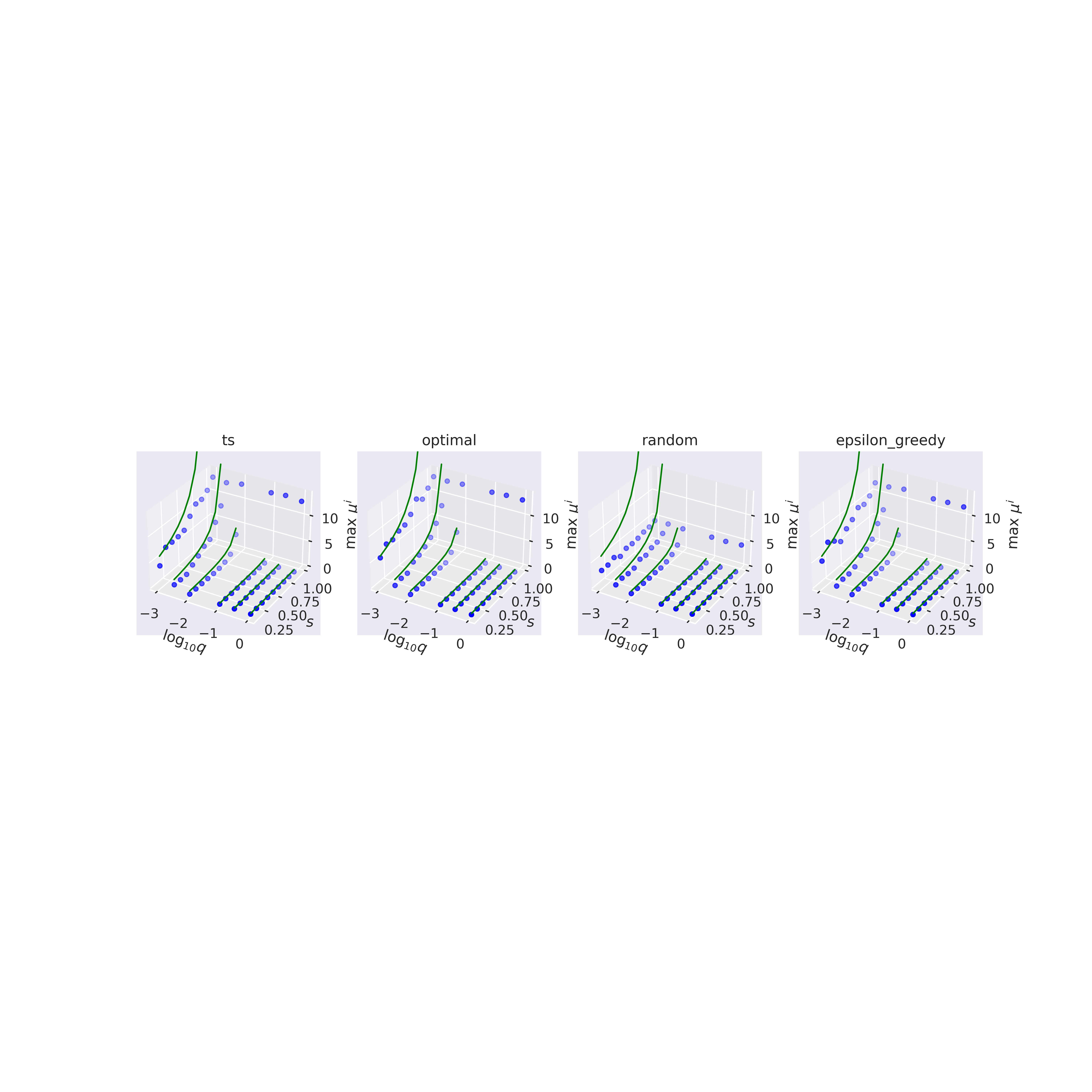}
    \caption{Maximum user interest averaged over 10 runs (blue dots) with interest restarts model for Thompson Sampling (ts), Random, Optimal and $\epsilon$-greedy (epsilon\_greedy) selection policies. Green line --- expected maximum user interest \eqref{eq:restarts-cond} for given scale and restart probability. Axes: y- scale parameter $s$, x- a logarithm of the restart probability $\log_{10} q$. Total number of steps $T=5000$.}
    \label{fig:restarts}
\end{figure}

We found that constant best lever assumption holds for Thompson Sampling (TS) $t$ gets large $t>1900$ and Optimal selection policies most of the time. While $\epsilon$-greedy ($\epsilon=0.1$) and Random policies do not exhibit the constant best lever property. At Fig.\,\ref{fig:add-noise-policies} we compare different selection policies for the additive noise model: Thompson Sampling (TS), Random, $\epsilon$-greedy, Optimal. The figure shows the user interest $\|\mu_t - \mu_0\|_2$ when $0 \leq t \leq 2000$. Results are averaged over and confidence intervals are shown for 30 runs. We can see that all policies exhibit a feedback loop when noise $w=3.0$. The interest grows much slower for the Random selection policy but the system still exhibits a feedback loop.  At Fig.\,\ref{fig:restarts} we show the amplitude of the feedback loop $\|\mu_t - \mu_0\|_2$ with respect to restart probability $q$ and scale $s$. We also plot the expected upper bound predicted by \ref{eq:restarts} on the same figure. It can be seen that at higher $q$ and $s$ the bound becomes tight. When the restart probability is low, non-optimality of the selection policy and insufficient number of steps $T=5000$ limits the growth of user interests. That is, the received reward is also bounded by $t \cdot \mathbf{E}\, \delta_t$.

With this we can confirm that Statement\,\ref{stmt:restarts} even when the constant best levers condition does not strictly hold. We also explore the interest restarts model for different values of available $M \in \{1,.., 10\}$ and selected $l \in \{1,.., 10\}$ levers and obtained similar results. Thus, RQ\,2 holds.

Results for different policies with the interest restarts model are also shown at Fig.\,\ref{fig:restarts}. Results are averaged over and confidence intervals are shown for 30 runs. As we can see, the predicted upper bound \ref{eq:restarts-cond} holds in all cases, although the Random policy demonstrates smaller growth in user interest than other policies, as expected. Considering the results we can confirm RQ\,3 that Statement\,\ref{stmt:add-noise} and Statement\,\ref{stmt:restarts} do not depend on the selection policy used.
\section{Analysis and limitations}
\label{sect:analysis}

In the experiment we find that additive noise does not prevent a feedback loop from occurring for several selection policies, and that interest restarts does limit the feedback loop. An upper bound is given by Statement 2 and it is tested in the experiments. Therefore, if an interest restart is possible in the system then the feedback loop is limited. The probability of restart and scale parameters should be estimated from the actual user behavior, which is a possible future direction of research. We find the constant best lever assumption useful for theoretical analysis as it greatly simplifies the proof. Results of the analysis are still valid when the assumption is violated in a feedback loop, as experiments show, because the feedback loop just reinforces the best lever already selected by the policy.

\paragraph{Limitations and validity.} We did not consider selection policies that expect data drift in user interests and, therefore, rewards. Non-stationary selection policies, such as Discounted Thompson Sampling (dTS) \cite{raj2017taming,besbes2014stochastic}, might be specifically suitable in case of sudden changes in user interests, especially for the interest restarts model. Our study is limited to theoretical models, which parameters need to be estimated from the actual user behavior. Nevertheless, our results still hold and useful, because we specify how predictions depend on the parameters even when assumptions are relaxed.

\section{Conclusions}

We state a problem of existence of feedback loops in presence of noise in user interests. We explore unbiased additive noise model and demonstrate that such type of noise does not affect the existence of the feedback loop both theoretically and experimentally. We also develop an interest restart model that models cases when users partly or completely lose interest recommended items. For this model we show that there exists an upper bound on the feedback loop if a restart is possible. We confirm our findings in the experiment. Further research could focus on parameter estimation and studying the non-stationary multi-armed problem.

%
%
%

\end{document}